\documentclass[journal,comsoc]{IEEEtran}
\usepackage[utf8]{inputenc}
\usepackage[english]{babel}

\usepackage{booktabs}
\usepackage{multirow}
\usepackage{hyperref}

\usepackage{cite}
\usepackage{graphicx}
\graphicspath{{figs/}}

\usepackage{amsmath}
\usepackage[cmintegrals]{newtxmath}

\usepackage{balance}

\usepackage{cancel}

\usepackage[table,xcdraw]{xcolor}
\usepackage{threeparttable, tablefootnote}

\usepackage{soul}
\soulregister\cite7
\soulregister\ref7
\soulregister\emph7
\soulregister\eqref7

\begin{document}

\title{Performance Analysis of Single-Cell\\Adaptive Data Rate-Enabled LoRaWAN}

\author{Arliones~Hoeller,~\IEEEmembership{Member,~IEEE,}
        Richard~Demo~Souza,~\IEEEmembership{Senior Member,~IEEE,}\\
        Samuel Montejo-Sánchez,~\IEEEmembership{Member,~IEEE,}
        and~Hirley~Alves,~\IEEEmembership{Member,~IEEE}\\\vspace{-.5cm}%
\thanks{A. Hoeller and R. D. Souza are with the Department of Electrical and Electronics Engineering of the Federal University of Santa Catarina, 88040-900 Florian{\'o}polis, Brazil.}%
\thanks{A. Hoeller and H. Alves are with the Centre for Wireless Communications of the University of Oulu, 90570 Oulu, Finland.}%
\thanks{A. Hoeller is also with the Department of Telecommunications Engineering of the Federal Institute for Education, Science, and Technology of Santa Catarina, 88130-310 S{\~a}o Jos{\'e}, Brazil.}%
\thanks{S. Montejo-Sanchez is with Programa Institucional de Fomento a la I+D+i, Universidad Tecnológica Metropolitana, 8940577 Santiago, Chile.}%
\thanks{Correspondence: Arliones.Hoeller@ifsc.edu.br, Richard.Demo@ufsc.br, SMontejo@utem.cl, Hirley.Alves@oulu.fi}%
\thanks{This work has been partially supported in Brazil by CNPq, FAPESC, project PrInt CAPES-UFSC ``Automation 4.0'', and INESC P\&D Brazil and Energisa (R\&D ANEEL PD-00405-1804/2018); in Finland by Academy of Finland (Aka) 6Genesis Flagship (Gr. 318927), EE-IoT (Gr. 319008), and Aka Prof (Gr. 307492); and in Chile by FONDECYT Postdoctoral (Gr. 3170021).}%
\thanks{\textcopyright 2020 IEEE. Personal use of this material is permitted. Permission from IEEE must be obtained for all other uses, in any current or future media, including reprinting/republishing this material for advertising or promotional purposes, creating new collective works, for resale or redistribution to servers or lists, or reuse of any copyrighted component of this work in other works.}}

\maketitle

\begin{abstract}
LoRaWAN enables massive connectivity for Internet-of-Things applications.
Many published works employ stochastic geometry to derive outage models of LoRaWAN over fading channels assuming fixed transmit power and distance-based spreading factor (SF) allocation.
However, in practice, LoRaWAN employs the Adaptive Data Rate (ADR) mechanism, which dynamically adjusts SF and transmit power of nodes based on channel state.
The community addressed the performance of ADR using simulations, but analytical models have not been introduced.
In this letter, we seek to close this gap.
We build over an analytical LoRaWAN model to consider the performance of steady-state ADR-enabled LoRaWAN.
We derive outage expressions and an optimization procedure to maximize the number of users under reliability constraints.
Results show that power allocation reduces interference and improves network capacity while reducing average power.
\end{abstract}

\vspace{-.3cm}
\section{Introduction}

Low-Power Wide-Area Networks (LPWAN) form a new class of technologies providing massive connectivity for the Internet-of-Things (IoT).
LPWAN technologies focus on Ma\-chi\-ne-Type Communications (MTC), especially on lightweight sensor network applications.
The most prominent LPWAN technologies are LoRaWAN, SigFox, and NB-IoT.
LoRaWAN has been widely used in academia due to openness and because it works in the unlicensed Industrial, Scientific, and Medical (ISM) bands~\cite{Centenaro:IEEEWC:2016}.
Several independent initiatives pushed the technology forward, making it available virtually everywhere.

Recent research on LoRaWAN shows that it may embrace the requirements of massive IoT applications.
Georgiou and Raza~\cite{Georgiou:WCL:2017} propose an analytic model of LoRaWAN disconnection and collision probabilities in Rayleigh fading channels. Disconnection considers the average probability that the signal-to-noise ratio (SNR) of a packet is below a reception threshold, while collision probability considers the threshold of the signal-to-interference ratio (SIR) of the same packet. The model captures the LoRaWAN sensitivity to collisions due to increased network usage, even though their SIR model only considers the dominant interferer.
Hoeller \textit{et al.}~\cite{Hoeller:Access:2019} extend~\cite{Georgiou:WCL:2017} and adapt the SIR model to consider several interference sources.
Mahmood \textit{et al.}~\cite{Mahmood:2019}, as well as \cite{Georgiou:WCL:2017} and \cite{Hoeller:Access:2019}, use stochastic geometry to build analytic coverage probability models for LoRaWAN and propose a path loss-based method to define network geometry.
Reynders \textit{et al.}~\cite{Reynders:ICC:2017} propose a power and data rate (spreading factor, SF) allocation method based on clustering for the NS-3 simulator.
Aligned to the problem we address, Abdelfadeel \textit{et al.}~\cite{Abdelfadeel:WoWMoM:2018} assess the performance of Adaptive Data Rate (ADR)-enabled LoRaWAN, achieving results similar to our theoretical analysis, and Li \textit{et al.} \cite{Li:Globecom:2018} study ADR convergence, both through simulations.

In this work, we review the analytic models for single-cell LoRaWAN and propose an adaptation to include the ADR feature.
Although multi-cell systems are likely to shape the topology of LoRaWAN networks in dense urban deployments, single-cell systems are still of interest for deployments in small town or villages, industrial plants, and in the agribusiness sector, where a dedicated single-cell LoRaWAN system may support a known number of users and applications.
Analytic models allow for faster evaluation and insights that are hard to obtain from simulations.
We validate our analytic model through Monte Carlo simulations.
Following \cite{Hoeller:Access:2019}, we use our model to plan the network deployment to respect a maximum outage probability.
We show that power control considerably reduces interference, increasing network capacity by up to $50\%$ and reducing average transmit power by roughly 25\%.

The main contributions in this letter are the performance analysis of ADR-enabled LoRaWAN and a simple closed expression for its outage probability in steady-state operation.
We assume the network reaches steady-state when ADR converges for all nodes, and their SF and transmit power configuration remain unchanged, as defined in \cite{Li:Globecom:2018}.
The performance analysis shows that ADR is an important feature of the technology and that it must be taken into account.
The closed-form expression assumes, as in \cite{Li:Globecom:2018}, that a network with static nodes converges to RSSI-based SF and transmit power figures, implementing, in practice, a truncated channel inversion scheme \cite{ElSawy:TWC:2014}.
Also, transient periods occur when channel or network conditions change, and the time to return to steady-state depends on application and deployment scenario~\cite{Li:Globecom:2018}.

\section{Baseline LoRaWAN Model}\label{sec:baseline}

LoRaWAN employs LoRa transceivers in the PHY layer, operating in sub-GHz frequencies (\textit{e.g.}, 868~MHz in Europe, 915~MHz in USA and Brazil) with Chirp Spread Spectrum modulation~\cite{Semtech:SX1276:2019}.
A key feature of LoRa modulation is the configurable SF rate.
As shown in Table~\ref{tab:lora}, higher SF rates increase signal robustness at the expense of transmission rate.
Since LoRa is a form of frequency modulation, it features the capture effect, where the receiver retrieves a colliding packet if it is sufficiently above interference.
The SIR for the successful reception of a packet is $6$dB~\cite{Semtech:SX1276:2019}.
A typical LoRa transceiver can use different transmit power ($\mathcal{P}$).
The Semtech SX1276 LoRa transceiver under European regulations admits 16 levels of transmit power between -1dBm and +14dBm, in 1dB steps.

\vspace{-.3cm}
\begin{table}[h]
\centering
\caption{LoRaWAN Uplink characteristics for packets of 19 bytes (13-bytes header, 6-bytes payload)~\cite{Semtech:SX1276:2019}.}
\label{tab:lora}
\scalebox{0.95}{
\begin{tabular}{@{}ccccc@{}}
\toprule
\textbf{\begin{tabular}[c]{@{}c@{}}SF\\ $i$\end{tabular}} & \textbf{\begin{tabular}[c]{@{}c@{}}ToA\\ $t_i$ (ms)\end{tabular}} & \textbf{\begin{tabular}[c]{@{}c@{}}Bitrate\\$Rb_i$ (kbps)\end{tabular}} & \textbf{\begin{tabular}[c]{@{}c@{}}Receiver Sensitivity\\ $\mathcal{S}_i$ (dBm)\end{tabular}} & \textbf{\begin{tabular}[c]{@{}c@{}}SNR threshold\\ $\psi_i$ (dB)\end{tabular}} \\ \midrule
7   &   51.46   & 5.46  & -123      & -6       \\
8   &  102.91   & 3.12  & -126      & -9       \\
9   &  185.34   & 1.75  & -129      & -12      \\
10  &  329.73   & 0.97  & -132      & -15      \\
11  &  741.38   & 0.53  & -134.5    & -17.5    \\
12  & 1318.91   & 0.29  & -137      & -20      \\ \bottomrule
\end{tabular}
}
\end{table}

In its most commonly used operating mode, known as class A, LoRAWAN implements a variation of unslotted ALOHA in a star network topology where nodes reach the gateway, which in turn connects to a network server via an IP network.

\vspace{-.3cm}
\subsection{Network Model}

\begin{figure}[tb]
    \centering
    \includegraphics[width=.7\columnwidth]{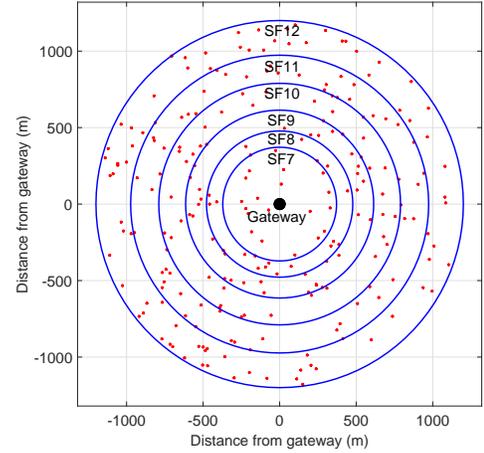}
    \caption{Sample of $\overline{N}=250$ nodes uniformly distributed in an area of radius $1200$m and with SF allocation for 1\% maximum disconnection probability.}
    \label{fig:nodes}
\end{figure}

We model the spatial distribution and activity of LoRaWAN nodes with stochastic geometry~\cite{Haenggi:Book:2012}.
We divide the network into SF rings according to the distance from the node to the gateway.
The vector $L=[l_0,\ldots,l_6], l_0=0,$ defines the SF ring edges, with $R=l_6$ as the coverage radius.
For simplicity, $S=\{1,\ldots,6\}$ is the set of SF rings, and each ring uses a respective SF in the set $\{7,\ldots,12\}$.
We consider that all nodes run the same application.
Thus network usage differs for each SF because of different data rates (see Time-on-Air/ToA in Table~\ref{tab:lora}).
We also assume that devices generate a packet for transmission once every $T$ seconds and that the packet is transmitted with a given probability according to the pure ALOHA protocol.
The transmission probability is a vector $p = [p_1, \ldots, p_6], p_i \in (0,1]~ \forall i \in S$, and $p_i = t_i/T$, where $t_i$ is the ToA of the packet with the SF of ring $i$.
For example, Figure~\ref{fig:nodes} presents a network configuration with $\overline{N}=250$ nodes and network geometry ($L$), obtained to ensure $0.99$ connection probability according to the method we describe in Section~\ref{sec:planning}.

Each SF ring constitutes a separate PPP $\Phi_i$ with intensity $\alpha_i=p_i\rho_i$ in its area $V_i = \pi (l_i^2 - l_{i-1}^2)$, where $l_{i-1}$ and $l_i$ form its inner and outer edges.
$\rho_i = \overline{N}_i/V_i$ is the spatial density of nodes in ring $i$.
The average number of nodes in $\Phi_i$ is $\overline{N}_i = \rho_i V_i$.
The average total number of nodes is $\overline{N} = \sum_{i \in S} \overline{N}_i$.
The coverage area is $V=\pi R^2$.
For instance, take ring $i=5$ (SF$_{11}$) in Figure~\ref{fig:nodes}, defined by two circles of radii $l_4=789.5$m and $l_5=973.4$m.
The ring area is $V_5=\pi(l_5^2 - l_4^2)=1.02$ km\textsuperscript{2}.
With $\overline{N}_5=\rho_5 V_5=50$ nodes in the ring, the spatial density is $\rho_5=\overline{N}_5/V_5=49.1$ nodes/km\textsuperscript{2}.
If the transmit probability is $p_5=0.01$, then the intensity of $\Phi_5$ is $\alpha_5=p_5\rho_5=0.49$.

In our analysis, $d_k$ is the Euclidean distance between the $k$-th node and the gateway, and $d_1$ denotes the distance of the node of interest to the gateway. We use the subscript ``1'' whenever a variable refers to the node under analysis.
Nodes use a transmit power $\mathcal{P}_k$ to send signal $s_k$, and both path loss and Rayleigh fading $h_k$ affect the signal $r_1$ received at the gateway.
Path loss follows $g_k = \left (\frac{\lambda}{4 \pi d_k}  \right )^\eta$, with wavelength $\lambda$, and path loss exponent $\eta>2$.
Therefore
\begin{align}
    r_1 &= s_1\sqrt{\mathcal{P}_1g_k}h_1 + \sum\nolimits_{k \in \Phi_i} {s_k\sqrt{\mathcal{P}_kg_k}h_k} + n,
\end{align}
where the first term is the attenuated signal of interest, the second is interference, $i$ is the ring of $s_1$, and $n$ is the zero-mean additive white Gaussian noise (AWGN) of variance $\mathcal{N}$.

\vspace{-.5cm}
\subsection{Outage Probability}

We consider that communication outage occurs due to disconnection or interference, which are, respectively, conditioned on the realization of the SNR and the SIR of a transmitted packet.
We base our analysis on the stochastic geometry model of the SINR of Poisson Bipolar Networks with Rayleigh fading in~\cite[Theorem 5.7]{Haenggi:Book:2012}.
Disconnection depends on distance and happens if the SNR is below the threshold $\psi_i$ (see Table~\ref{tab:lora}).
The disconnection probability is~\cite{Georgiou:WCL:2017}
\begin{align}
    H_0(d_1,\mathcal{P}_1) = \mathbb{P}[\textup{SNR} < \psi_i] = \mathbb{P} \left[ \frac{\mathcal{P}_1 g_1 |h_1|^2}{\mathcal{N}} < \psi_i ~\biggr|~ d_1 \right], \nonumber
\end{align}
with $i$ indicating the SF ring in use by the node under analysis.
With known $d_1$ and $\mathcal{P}_1$, we condition $H_0$ to the probability of the Rayleigh fading power in $|h_1|^2\sim\exp(1)$, so
\begin{align}
    H_0(d_1,\mathcal{P}_1) &= 1 - \textup{exp}\left( - \frac{\psi_i\mathcal{N}}{\mathcal{P}_1 g_1} \right). \label{eqn:h0}
\end{align}

The outage due to interference ({\it i.e.}, collision with other packets) considers the capture effect.
Thus, the collision probability concerning the SIR threshold $\delta$ is~\cite{Hoeller:Access:2019}
\begin{align}
    Q_0(d_1,\mathcal{P}_1) \!=\! \mathbb{P}\![\textup{SIR} \!<\! \delta | d_1]
    \!=\! \mathbb{P} \left[ \frac{\mathcal{P}_1 g_1 |h_1|^2}{\sum_{k\in\Phi_i} \mathcal{P}_k g_k |h_k|^2} \!<\! \delta \biggr| d_1 \right]. \label{eqn:q0}
\end{align}

\section{Power Allocation for LoRaWAN}\label{sec:model}

When considering transmit power allocation, $\mathcal{P}_k$ may be different for each node.
We assume that nodes at the edge of each SF ring use the highest available transmit power ($\mathcal{P}_{max}$) to extend the coverage area.
Considering a predefined target outage due to disconnection ($\mathcal{T}_{H_0}$), we define the network geometry by making $H_0(l_i,\mathcal{P}_{max}) = \mathcal{T}_{H_0}$, so that 
\begin{align}
    l_i &= \frac{\lambda}{4\pi} \left( - \frac{\mathcal{P}_{max}\textup{ln}(1-\mathcal{T}_{H_0})}{\mathcal{N}\psi_i} \right)^{\frac{1}{\eta}}. \label{eqn:l_i}
\end{align}

We also use~\eqref{eqn:h0} to define the minimum transmit power the $k$-th device must use to ensure $\mathcal{T}_{H_0}$ as
\begin{align}
    \mathcal{P}_{k_{min}} &= - \frac{\mathcal{N}\psi_i}{\textup{ln}(1-\mathcal{T}_{H_0}) g_k}. \label{eqn:pkmin}
\end{align}
In practice, $\mathcal{P}_{k_{min}}$ should be rounded up to the immediately higher value available.
Additionally, we obtain the network average transmit power by averaging~\eqref{eqn:pkmin} over the area, \textit{i.e.},
\begin{align}
    \mathcal{P}_{avg} &= \frac{2\pi}{V} \sum_{i\in S} \int_{l_{i-1}}^{l_i} -\frac{\mathcal{N}\psi_i}{\textup{ln}(1-\mathcal{T}_{H_0}) g_k} d_k ~\textup{d}d_k \nonumber \\
    &= -\frac{2\pi\mathcal{N}}{V\textup{ln}(1-\mathcal{T}_{H_0})} \left( \frac{4\pi}{\lambda} \right)^\eta \sum_{i\in S} \frac{\psi_i}{\eta+2} (l_i^{\eta+2} - l_{i-1}^{\eta+2}). \label{eqn:pavg}
\end{align}

\subsection{Outage Probability with Transmit Power Allocation}

Rewriting the disconnection probability in~\eqref{eqn:h0} with the power allocation method defined by~\eqref{eqn:pkmin} yields
\begin{align}\label{eqn:h1final}
    H_0(d_1,\mathcal{P}_{1_{min}}) = 1 - \textup{exp}\left( - \frac{q_1\mathcal{N}}{\mathcal{P}_{1_{min}} g_1} \right) = \mathcal{T}_{H_0},
\end{align}so that transmit power control compensates for path loss, makes $H_0$ independent of $\mathcal{P}_1$ and $d_1$, and ensures $\mathcal{T}_{H_0}$ for all nodes.
Similarly, rewriting~\eqref{eqn:q0} with~\eqref{eqn:pkmin} yields
\begin{align}
   Q_0(i) &= \mathbb{P} \left[ \frac{|h_1|^2}{\sum_{k\in\Phi_i} |h_k|^2} < \delta \right],
\end{align}
and therefore $Q_0$ becomes independent of transmit powers and distances from the gateway, being only dependent on fading.

If we define $X_i = \sum_{k\in\Phi_i} |h_k|^2$ and $Y_i = \frac{|h_1|^2}{X_i}$, then $Q_0(i) = \mathbb{P}\left[ Y_i < \delta \right] = F_{Y_i}(\delta)$, with the cdf of $Y_i$ obtained as
\begin{align}
    F_{Y_i}(y) = \int_0^\infty F_{|h_1|^2}(xy) f_{X_i}(x)~\textup{d}x,\label{eqn:Fh1X}
\end{align}
where $|h_1|^2 \sim \textup{exp}(1)$, $F_{|h_1|^2}(z) = 1 - e^{-z}$, $X_i$ is Gamma distributed, $X_i\sim\Gamma(N_{\Phi_i},1)$, $f_{X_i}(x)=\frac{1}{\Gamma(N_{\Phi_i})}x^{N_{\Phi_i}-1}e^{-x}$, and $\Gamma(\cdot)$ is the Gamma Function~\cite{NIST:Book:2010}.
Following the duality of notation of PPPs~\cite[Box 2.3]{Haenggi:Book:2012}, $N_{\Phi_i}\sim\textup{Poiss}(\beta_i)$ is a Poisson random variable of mean $\beta_i=\alpha_iV_i=p_i\overline{N}_i$ describing the average number of \emph{active} interferers in PPP $\Phi_i$.
Thus,
\begin{align}
    Q_0(i) = \mathbb{E}_{N_{\Phi_i}} \left[ \int_0^\infty (1 - e^{-x\delta}) \frac{1}{\Gamma(N_{\Phi_i})} x^{N_{\Phi_i}-1} e^{-x} \textup{d}x \right],
\end{align}
which is solved by distributing the multiplication, factoring out independent terms, and applying the identity $\int_0^\infty x^n e^{-ax} \textup{d}x = \frac{\Gamma(n+1)}{a^{n+1}}$~\cite{NIST:Book:2010}. Thus, the $N_{\Phi_i}$-dependent collision probability is
\begin{align}
    Q_0(i) = \mathbb{E}_{N_{\Phi_i}} \left[ 1 - (\delta + 1)^{-N_{\Phi_i}} \right].
\end{align}
Since the pmf of $N_{\Phi_i}$ is $f_{N_{\Phi_i}}(z) = \frac{\beta_i^z e^{-\beta_i}}{z!}$,
\begin{align}
    Q_0(i) = 1 - \textup{exp}\left( - \frac{\delta}{\delta + 1}\beta_i \right). \label{eqn:q0final}
\end{align}

Finally, the total outage probability for each SF ring $i$ is
\begin{align} 
    C_0(i) \triangleq H_0 + Q_0(i) - H_0Q_0(i). \label{eqn:c0}
\end{align}
Our model preserves the PPP properties for each point as long as the fixed communication distances and transmit powers sa\-tis\-fy $\frac{\mathcal{P}_1g_1}{\mathcal{P}_kg_k}=1$ in~(3), which is guaranteed by~\eqref{eqn:pkmin}.

\section{Network planning}\label{sec:planning}

We use the outage probability in~\eqref{eqn:c0} as a tool to plan the deployment of single-cell LoRaWANs. We assume a target maximum outage $\mathcal{T}_{C_0}$ for all nodes, $C_0(i) \leq \mathcal{T}_{C_0}, \forall i$.
We use this reliability constraint to maximize coverage radius and network usage.
After a closer look at~\eqref{eqn:c0}, we observe that, for each ring, $C_0(i)$ depends on the outer limit $l_i$ and the average number of active interferers $\beta_i$. Unfortunately, it is not possible to solve such optimization for both variables simultaneously, so, here, we explore the trade-off between coverage radius and network usage.
Assuming that the larger coverage radius and higher network usage occur on the worst-case scenario where $C_0(i)=\mathcal{T}_{C_0}, \forall i$, we represent the trade-off, following from~\eqref{eqn:c0}, as $\mathcal{T}_{C_0} = \mathcal{T}_{H_0} + Q_0(i) - \mathcal{T}_{H_0} Q_0(i)$, from which we equate, either, the maximum $\beta_i$ assuming a given $\mathcal{T}_{H_0}$ as
\begin{align}
    \beta_i = -\frac{\delta + 1}{\delta} \textup{ln} \left( \frac{1-\mathcal{T}_{C_0}}{1-\mathcal{T}_{H_0}} \right), \label{eqn:bi}
\end{align}
or the maximum $\mathcal{T}_{H_0}$ assuming a given $\beta_i$ as
\begin{align}
    \mathcal{T}_{H_0} = \frac{\mathcal{T}_{C_0} - Q_0(i)}{1 - Q_0(i)}. \label{eqn:th0}
\end{align}

Note that $\beta_i=p_iN_i$, so we use~\eqref{eqn:bi} to obtain the maximum number of nodes in each ring, assuming that all nodes in a ring use the same duty-cycle $p_i$.
Similarly, because of~\eqref{eqn:l_i}, we obtain the SF ring range $l_i$ with $\mathcal{T}_{H_0}$ from~\eqref{eqn:th0}.

\section{Numerical Results}\label{sec:results}

\begin{table}[tb]
\centering
\caption{Model and simulation parameters.} \label{tab:param}
\scalebox{.95}{
\begin{tabular}{@{}ll@{}}
\toprule
\textbf{Parameter}            & \textbf{Value   }                    \\ \midrule
$f_c$                & 868 MHz                     \\
$B$                  & 125 kHz                     \\
$NF$                 & 6 dB                        \\
$\mathcal{N}$        & $-174 + NF + 10 \textup{log}_{10}(B) = -117$dBm \\
$T$                  & Every 15 minutes            \\
$p~(\times 10^{-6})$ & $\{57.1, 114.3, 205.9, 366.3, 823.7, 1465.4 \}$ \\
$\mathcal{P}_k$      & $\{-1, 0, \ldots, 14\}$ dBm \\
$\mathcal{P}_{avg}$  & 12.63 dBm                   \\
$\mathcal{P}_{max}$  & 14 dBm                      \\
$\delta$             & 6 dB                        \\
$\mathcal{T}_{C_0}$  & 0.01                        \\
$R_{min}$            & 1200 m                      \\ \bottomrule
\end{tabular}
}
\end{table}

\begin{figure}[tb]
    \centering
    \includegraphics[width=.94\columnwidth]{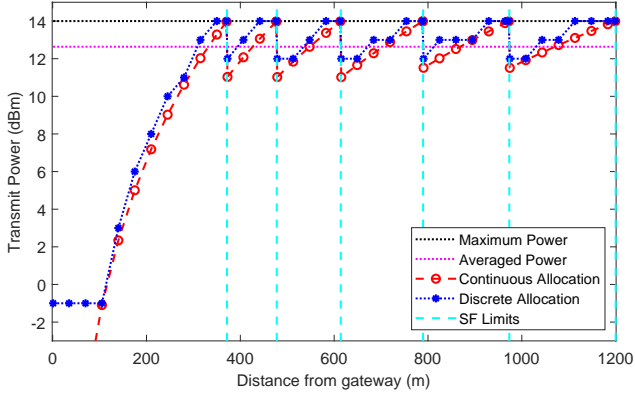}
    \caption{Power allocation as a function of distance.}
    \label{fig:ptx_alloc}
\end{figure}

We assume the parameters in Tables~\ref{tab:lora} and~\ref{tab:param} to mimic a suburban deployment of a single-cell LoRaWAN under European regulations.
The figures show our theoretical model (solid lines) and Monte Carlo simulations (marks).
Figure~\ref{fig:ptx_alloc} shows the power allocation using~\eqref{eqn:pkmin} and the average power in the network.
The dashed curve shows the continuous power allocation according to distance and considering different SFs. It shows that $SF_7$ uses a wider range of transmit power because its nodes are closer to the gateway. The power variation is 3dB in $SF_8$, $SF_9$, and $SF_{10}$, and 2.5dB in $SF_{11}$ and $SF_{12}$. That matches the variation of the SNR threshold in Table~\ref{tab:lora} ($\psi_i$) and is also aligned with the ADR power and SF allocation method defined by LoRaWAN.
Still, in Figure~\ref{fig:ptx_alloc}, the dotted curve shows the discrete practical power allocation, obtained by rounding up the continuous values of~\eqref{eqn:pkmin}. 
That mostly impacts the power of nodes closer to the gateway. Figure~\ref{fig:ptx_alloc} also shows the average power in the network from~\eqref{eqn:pavg} as $12.63$~dBm -- an average power reduction of $27\%$.

\begin{figure}[tb]
    \centering
        \includegraphics[width=.98\columnwidth]{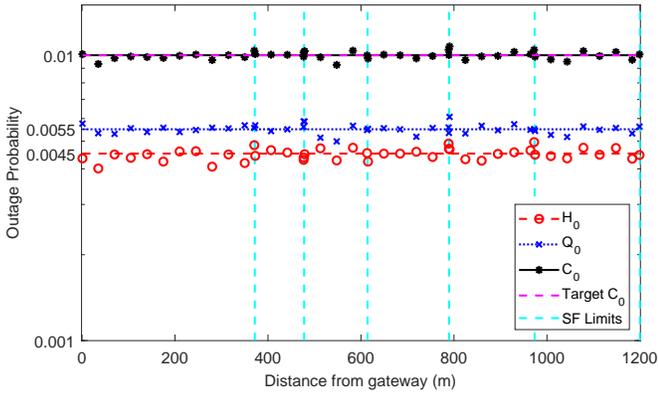}
        \caption{System performance with power allocation. $\overline{N}=247$.}
        \label{fig:disc_ptx}
\end{figure}

\begin{figure}[tb]
        \centering
        \includegraphics[width=.98\columnwidth]{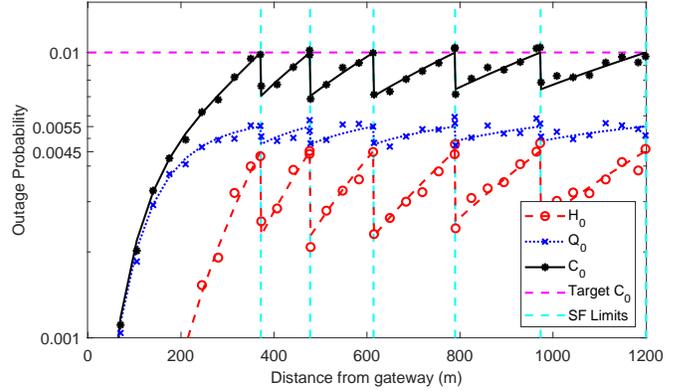}
               \caption{System performance with fixed power. $\overline{N}=225$.}
                \label{fig:full_ptx}
\end{figure}

Figures~\ref{fig:disc_ptx} and~\ref{fig:full_ptx} show results using two approaches: power allocation as in~\eqref{eqn:pkmin}, and all nodes with maximum power ($14$~dBm). The most noticeable aspect is that proper power allocation allows all nodes in the network to experience similar outage probabilities close to the target $\mathcal{T}_{C_0}=0.01$. When nodes use constant power, $\mathcal{T}_{C_0}$ is reached only on the edges of each SF ring. In the constant power scenario, the nodes closer to the ring inner edge use more power than needed, thus spending more energy and causing more interference. 

The method in Figure~\ref{fig:disc_ptx}, besides using less average power than that in Figure~\ref{fig:full_ptx}, also serves more users. We observe a gain of 9.3\% in the number of supported users, on average, from 225 to 247 nodes. If we consider a scenario with fixed transmit power equal to the average power used in Figure~\ref{fig:disc_ptx}, then power allocation leads to a gain of 56.7\% in the number of users, from 157 to 247. Our results show that adequate power allocation in LoRaWAN contributes to the network capacity due to the interference reduction while being more energy-efficient.


\section{Conclusion}\label{sec:conclusion}

We modeled the performance of LoRaWAN with power allocation, considering two outage conditions: disconnection and interference. We determined the maximum number of users to ensure a maximum outage probability. Numerical results show that power allocation increases network reliability due to the reduction of interference while being more energy-efficient than fixed transmit power. In the future, we plan to investigate the performance of LoRaWAN with power control under inter-SF and external interference.


\bibliographystyle{IEEEtran}
\bibliography{hoeller_WCL1630}

\begin{thebibliography}{10}
\providecommand{\url}[1]{#1}
\csname url@samestyle\endcsname
\providecommand{\newblock}{\relax}
\providecommand{\bibinfo}[2]{#2}
\providecommand{\BIBentrySTDinterwordspacing}{\spaceskip=0pt\relax}
\providecommand{\BIBentryALTinterwordstretchfactor}{4}
\providecommand{\BIBentryALTinterwordspacing}{\spaceskip=\fontdimen2\font plus
\BIBentryALTinterwordstretchfactor\fontdimen3\font minus
  \fontdimen4\font\relax}
\providecommand{\BIBforeignlanguage}[2]{{%
\expandafter\ifx\csname l@#1\endcsname\relax
\typeout{** WARNING: IEEEtran.bst: No hyphenation pattern has been}%
\typeout{** loaded for the language `#1'. Using the pattern for}%
\typeout{** the default language instead.}%
\else
\language=\csname l@#1\endcsname
\fi
#2}}
\providecommand{\BIBdecl}{\relax}
\BIBdecl

\bibitem{Centenaro:IEEEWC:2016}
{M. Centenaro \textit{et al.}}, ``Long-range communications in unlicensed
  bands: the rising stars in the {IoT} and smart city scenarios,'' \emph{IEEE
  Wireless Commun.}, vol.~23, no.~5, pp. 60--67, Oct 2016.

\bibitem{Georgiou:WCL:2017}
O.~Georgiou and U.~Raza, ``Low power wide area network analysis: Can {LoRa}
  scale?'' \emph{IEEE Wireless Commun. Lett.}, vol.~6, no.~2, pp. 162--165, Apr
  2017.

\bibitem{Hoeller:Access:2019}
{A. Hoeller \textit{et al.}}, ``Optimum {LoRaWAN} configuration under {Wi-SUN}
  interference,'' \emph{IEEE Access}, vol.~7, pp. 170\,936--170\,948, Dec 2019.

\bibitem{Mahmood:2019}
{A. Mahmood \textit{et al.}}, ``Scalability analysis of a {LoRa} network under
  imperfect orthogonality,'' \emph{IEEE Trans. Ind. Informat.}, vol.~15, no.~3,
  pp. 1425--1436, Mar 2019.

\bibitem{Reynders:ICC:2017}
{B. Reynders \textit{et al.}}, ``Power and spreading factor control in low
  power wide area networks,'' in \emph{Proc. IEEE International Conference on
  Communications}, May 2017, pp. 1--5.

\bibitem{Abdelfadeel:WoWMoM:2018}
{K. Abdelfadeel \textit{et al.}}, ``Fair adaptive data rate allocation and
  power control in {LoRaWAN},'' in \emph{Proc. IEEE International Symposium on
  a World of Wireless, Mobile and Multimedia Netw.}, Jun 2018, pp. 1--9.

\bibitem{Li:Globecom:2018}
{S. Li \textit{et al.}}, ``How agile is the adaptive data rate mechanism of
  {LoRaWAN}?'' in \emph{Proc. IEEE Global Communications Conference}, Dec 2018,
  pp. 206--212.

\bibitem{ElSawy:TWC:2014}
H.~{ElSawy} and E.~{Hossain}, ``On stochastic geometry modeling of cellular
  uplink transmission with truncated channel inversion power control,''
  \emph{IEEE Trans. Wireless Commun.}, vol.~13, no.~8, pp. 4454--4469, Aug
  2014.

\bibitem{Semtech:SX1276:2019}
\emph{SX1276/77/78/79 - 137 {MHz} to 1020 {MHz} Low Power Long Range
  Transceiver}, Semtech Coorporation, Jan 2019.

\bibitem{Haenggi:Book:2012}
M.~Haenggi, \emph{Stochastic Geometry for Wireless Networks}.\hskip 1em plus
  0.5em minus 0.4em\relax Cambridge: Cambridge University Press, 2012.

\bibitem{NIST:Book:2010}
{Olver \textit{et al.}}, \emph{NIST Handbook of Mathematical Functions}.\hskip
  1em plus 0.5em minus 0.4em\relax Cambridge: Cambridge University Press, 2010.

\end{thebibliography}

\end{document}